\documentstyle[11pt,newpasp,epsf]{article}

\begin{document}

\title{Dust grain properties in atmospheres of AGB stars}

\author{A.\,C.~Andersen} 
\affil{Niels Bohr Institute for Astronomy, Geophysics and Physics, Copenhagen University, Juliane Maries Vej 30, DK-2100 Copenhagen, Denmark}

\author{S.~H\"ofner}
\affil{Department of Astronomy and Space Physics, Uppsala University, 
Box 515, SE-751\,20 Uppsala, Sweden}

\author{R.~Gautschy-Loidl}
\affil{R\"uti, Switzerland} 






\begin{abstract}
We present self-consistent dynamical models for dust driven winds
of carbon-rich AGB stars. The models are based on the coupled system of frequency-dependent 
radiation hydrodynamics and time-dependent dust formation. We investigate in detail
how the wind properties of the models are influenced by the micro-physical properties of the
dust grains that enter as parameters. The models are now at a level where it is
necessary to be quantitatively consistent when choosing the dust properties that enters as input into
the models. At our current level of sophistication the choice of dust parameters is 
significant for the derived outflow velocity, the degree of condensation and the 
estimated mass loss rates of the models.
In the transition between models with and without mass-loss the choice of
micro-physical parameters turns out to be very significant for whether a particular set of stellar
parameters will give rise to a dust-driven mass loss or not.
\end{abstract}




%
%
%
\section{Introduction}

Mass loss by dust driven winds of asymptotic giant branch (AGB) stars
probably is one of the major mechanism which recycle
material in the Galaxy (e.g.\ Sedlmayr 1994). Most stars
($M_{\star} < 8 M_{\odot}$) will eventually become AGB stars
and subsequently end their life as white dwarfs surrounded by 
planetary nebulae.

AGB stars are cool ($T_{\star} < 3500$~K) and luminous ($L_{\star}$ of a few
$10^{3}$ to a few $10^{4}$~L$_{\odot}$)
and a majority of them are pulsating long-period variables (LPVs). 
The outer layers of many AGB stars provide favorable conditions for 
the formation of molecules and dust grains. Dust grains play an
important role for the heavy mass loss of these stars (up to 
($\dot{M} \sim 10^{-4} M_{\odot}$/yr). 

Pulsation causes an extended atmosphere where the dust is
condensing. The dust absorbs the
light of the central star and re-radiates it at longer wavelengths in the
infrared rage ($\lambda > 2 \mu$m). The density of material in the
circumstellar envelope may be so large that the star becomes completely obscured 
in the optical rage.
Due to observational difficulties in determining the mass loss
rates of these objects considerable effort has been put into
a theoretical description of the mass loss of AGB stars (e.g.\ 
Fleischer et al.\ 1992; H\"ofner \& Dorfi 1997).  The overall goal is to develop 
a mass loss description that can be used as input to
evolutionary models and thereby probe the chemical evolution of the
Galaxy.

We present here how the wind properties of carbon-rich models are 
influenced by the micro-physical properties of the dust grains. 

\begin{figure}
\plotfiddle{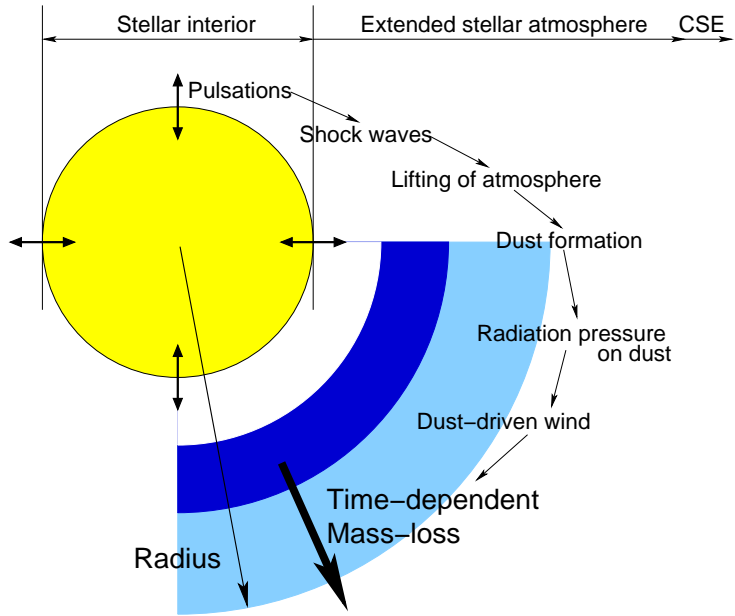}{10 cm}{0}{45}{45}{-150}{-0} 
\caption{A schematic drawing of the atmosphere of an AGB star and the
most important physical processes. (Courtesy of Christer Sandin).}
\end{figure}

\section{The Atmosphere Model}  

The models predict mass loss rates, wind velocities and dust-to-gas ratios 
by solving the equations
of continuity, motion and energy for the gas and a set of equations describing the
formation of dust grains. The transfer of radiation is calculated 
frequency-dependently for the gas and dust (H\"ofner 1999; H\"ofner et al.\ 2002).
Basically only four
stellar parameters are needed as input for the models; the 
stellar mass ($M_{\star}$),
the effective temperature ($T_{\rm eff}$), the luminosity
($L_{\star}$) of the star and the gas abundance ratio
(${\varepsilon_{\rm C}/\varepsilon_{\rm O}}$) of carbon
to oxygen. However, to describe the pulsations of the star two additional
parameters are required for the prescription of the inner boundary
condition; a pulsation period (P) and a velocity amplitude ($\Delta v$),
which simulates a pulsating stellar surface as a means of mechanical 
energy input. 

The number densities of the molecules relevant
to the carbon dust formation are calculated assuming chemical equilibrium
between H, H$_{2}$, C, C$_{2}$, C$_{2}$H and C$_{2}$H$_{2}$ after
the fraction of carbon bound in CO has been subtracted. Nucleation,
growth and destruction of dust grains are supposed to proceed
by reactions involving C, C$_{2}$, C$_{2}$H and C$_{2}$H$_{2}$.

Dust is formed by a series of chemical reactions in which atoms or
molecules from the gas phase combine to clusters of increasing
size. The molecular composition of the gas phase
determines which atoms and molecules are available for the cluster
formation and grain growth. This can lead to a layered structure of 
the circumstellar envelope. Dust formation alters the atmospheric structure which  
significant influence the local and global circumstellar envelope dynamics and
the detailed time-dependent spectral appearance. 

In the case of pulsating atmospheres of Mira and other LPVs the 
condensation of dust grains is strongly connected with the shock 
wave structure of the circumstellar envelope, usually triggered by a 2-step mechanism.  
First there is  nucleation of critical clusters followed by growth to 
macroscopic dust grains.  Grain growth will
usually proceed far from equilibrium and it is necessary to use a time-dependent description of grain growth to determine the degree of 
condensation and other relevant properties of the dust.

The grain growth is treated by the moment method which describe the
time evolution of an ensemble of macroscopic dust grains of various sizes and
requires the nucleation rate as input
(Gail \& Sedlmayr 1988; Gauger et al.\ 1990).
The moment method needs the 
following dust properties as external input: 
(1) the optical properties of the relevant dust types,
(2) the intrinsic density of the dust material(s),
(3) the sticking coefficient which describes the efficiency of
growing dust grains from the atomic precursors, and
(4) the surface tension of the dust grains which controls the
efficiency of nucleation. 
The purpose of this work is to investigate in detail how the predicted 
wind properties of the carbon-rich AGB models are influenced by the 
choice of these quantities.

For the mass loss models it is important to know
how the uncertainty in the chosen dust parameters 
affects the obtained results. The model parameters are listed in Tables $2-5$.
For a given set of stellar parameters and elemental abundance, the
mass loss rate, the outflow velocity and the degree of condensation
is obtained by averaging the outflow of the time-dependent models.
Previous models presented in earlier papers (e.g.\ Andersen et al.\ 1999)  
were all based
on gray radiative transfer while the models presented here are
calculated using frequency dependent radiative transfer for the
gas and dust (see H\"ofner et al.\ 2002 for details).
Dynamic atmosphere models are undergoing a transition from qualitative
modeling of physical processes to realistic quantitative predictions
and interpretations of observations.

\begin{figure}
\plotfiddle{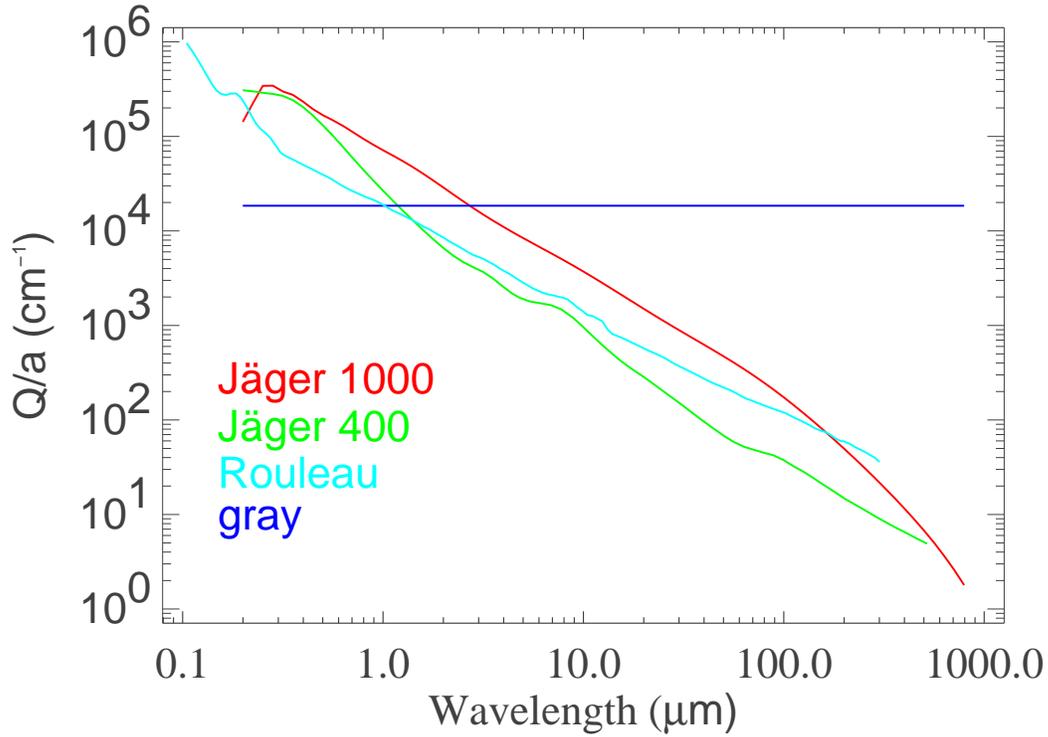}{10 cm}{0}{85}{85}{-270}{-320} 
\caption{Wavelength dependence of the dust extinction efficiency $(Q_{\rm ext}/a)$,
           see Table \ref{opacitydust} for dust annotation. Also shown is 
           a constant value (gray opacity) normalised to the {\it Rouleau} data at $1 \mu$m.}
\label{gray}
\end{figure}

\section{Amorphous carbon dust}

\begin{table}
\caption{List of the different dust data used.}
\begin{center}
\begin{tabular}{|l|c|c|c|c|} \hline
Reference & Material & $\rho_{\rm dust}$ & $sp^2$ & Designation  \\
 & name & (g/cm$^{3}$) & \% & in this paper  \\ \hline
J\"{a}ger et al.\ 1998 & cel400  & 1.435 & 67 & {\it J\"{a}ger~400}  \\
J\"{a}ger et al.\ 1998 & cel1000 & 1.988 & 80 & {\it J\"{a}ger~1000}  \\
Rouleau \& Martin 1991 & AC2 & 1.85 & -  & {\it Rouleau} \\ \hline
\end{tabular}
\end{center}
\label{opacitydust}
\end{table}

Amorphous carbon particles are considered to be a very good candidate for the most common type of dust
present in circumstellar envelopes of carbon-rich AGB stars. The infrared data of
late-type stars generally show a dust emissivity law ($Q(\lambda) \sim \lambda^{- \beta}$) with
a spectral index of $\beta \sim 1$ (e.g. Campbell et al.\ 1976;
Sopka et al.\ 1985; Martin \& Rogers 1987;
G\"urtler et al.\ 1996).  A $\lambda^{-1}$ behavior can be expected in a very
disordered two-dimensional material like amorphous material (e.g. Huffmann 
1988; J\"ager et al.\ 1998). 
Graphite formation in AGB stars is ruled out, because of
the absence in the observed spectra of the narrow band at $11.52 \mu$m and the overall shape
of infrared graphite spectra which are  proportional to
$\lambda^{-2}$ (e.g. Draine \& Lee 1984).

Amorphous carbon covers by nature a broad span of material properties. The
difference in optical properties for three different types of amorphous carbon
can be seen in Fig.\,1. The dust properties needed as input for the models are
not always know from laboratory experiments. In the case of amorphous carbon it
has therefore been custom to use values of the intrinsic density 
for graphite material instead of 
the real value. If the intrinsic density of graphite is used instead of the 
correct laboratory value for amorphous carbon it will influence the efficiency of
dust formation and therefore the predicted mass loss and
colors as seen from Table 3. 

\section{Results and Discussion}  

A frequency-dependent treatment of the radiative transfer in the
dynamical models has turned out to be crucial for obtaining realistic
structures, synthetic spectra, and reliable mass loss rates (H\"ofner et al.\ 2002). 
How the choice of the micro-physical 
parameters can change the predictions for the mean outflow velocity of 
the gas and dust, the degree of dust condensation and the 
mass loss is discussed in the following focusing on the
optical dust properties, the intrinsic dust density, the 
sticking coefficient and the surface tension.  

\subsection{Dust opacity} 

The direct influence of the extinction efficiency ($Q_{\rm ext}$) on the structure
and the wind of the dynamical models are shown in Table\,2.
The models are dependent on the choice of laboratory measurements of
the extinction efficiency ($Q_{\rm ext}$) as already show
for gray models in Andersen et al.\ (1999).
In contrast to the gray models then for the frequency-dependent models
both the absolute value and the slope of the dust opacity data become
relevant. The absolute value of the extinction efficiency 
has a profound influence on the determined outflow velocity of
the models, while the slope of the extinction efficiency determines 
the grain temperature.
In the models where the {\it J\"ager~400} data is used for the extinction efficiency
of the dust, no dust is formed, since the slope of the
material opacity dictates a high grain temperature which make the grains 
unstable against evaporation.

\begin{table}
\caption{Influence of the extinction efficiency of the dust. 
         {\it Model parameters:} $M_{\star} = 1.0 \,M_{\odot}$,
         sticking coefficients $\alpha_{C} = 0.37$ $\alpha_{C_{2}} = 0.34$ 
         $\alpha_{C_{2}H} = 0.34$ $\alpha_{C_{2}H_{2}} = 0.34$,
         abundances are solar except for C/O = 1.4,
         period $P = 650$\,d, piston velocity ${\Delta u_{\rm p}} = 4$\,km/s.  
         {\it Input:} Luminosity $L_{\star}$, temperature $T_{\star}$,
         dust opacity data $\kappa_{\rm dust}$, 
         intrinsic dust density $\rho_{\rm dust}$. 
         {\it Output:} Mass loss rate $\dot{M}$,
         mean velocity at the outer boundary ${\langle u \rangle}$,
 mean degree of condensation at the outer boundary ${\langle f_{\rm c} \rangle}$.} 
\begin{center}
\begin{tabular}{|l|c|c|c|c||c|c|c|}
 \hline
Model& $L_{\star}$ & $T_{\star}$  & $\kappa_{\rm dust}$& $\rho_{\rm dust}$ & $\dot{M}$ & ${\langle u \rangle}$ & ${\langle {f_{\rm c}} \rangle}$  \\
 & [$L_{\odot}$] & [K] & & [g/cm$^{3}$] & [${\rm {M_{\odot}} / yr}$] & [km/s] &  \\ \hline
l13dj10$\rho$199&  13000 & 2700 & {\it J\"ager~1000} & 1.99 & $5.59 \cdot 10^{-6}$ & 14.8 & 0.05  \\
l13drou$\rho$185& 13000 & 2700 & {\it Rouleau} & 1.85 &  $4.87 \cdot 10^{-6}$ & 7.38 & 0.10 \\
l13dj04$\rho$144& 13000 & 2700 & {\it J\"ager~400} & 1.44 &  - & - & -  \\
l10dj10$\rho$199& 10000 & 2600 & {\it J\"ager~1000} & 1.99 &  $7.0 \cdot 10^{-6}$ & 16.3 & 0.10  \\
l10drou$\rho$185&  10000 & 2600 & {\it Rouleau} & 1.85 &  $2.26 \cdot 10^{-6}$ & 3.64 & 0.12  \\
l10dj04$\rho$144&  10000 & 2600 & {\it J\"ager~400} & 1.44 &   - & - & -  \\
\hline
\end{tabular}
\end{center}
\end{table}

   \begin{figure}
\plotfiddle{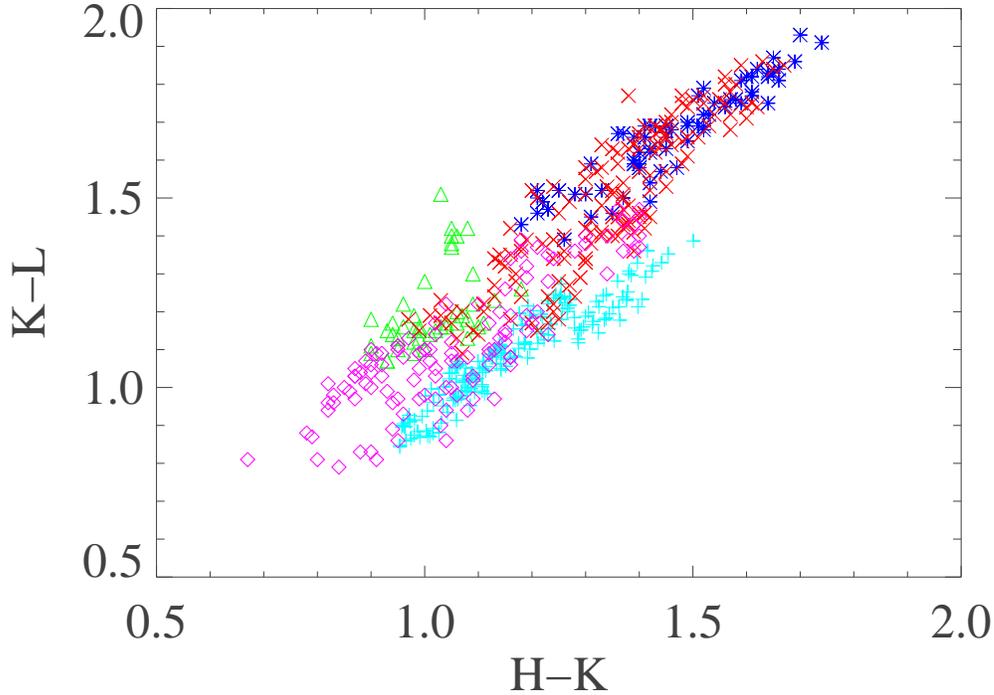}{8 cm}{0}{85}{85}{-280}{-335} 
     \caption{The colors ($H-K$) and ($K-L$) for the models ($+$) 
  presented in Tables $2-5$ and for carbon Mira with a comparable 
  mass loss rate;  V Hya (${\small \triangle}$),
  R Vol ($\ast$), R For ($\times$), R Lep (${\small \diamond}$).}
         \label{massloss_color}
   \end{figure}

\subsection{Intrinsic dust density} 

We have calculated models with 
different intrinsic densities for the given material
assumed to be formed in the stellar wind. When the moment method was
developed by Gail \& Sedlmayr (1988) the intrinsic
density for the amorphous carbon used in that generation of models
was not know. The value of graphite (2.25 g/cm$^{3}$)
was therefore assumed. This value was later used in almost all of
the existing models (e.g.\ Fleischer et al.\ 1992; H\"ofner \& Dorfi 1997).
The models become much redder when the intrinsic density of graphite
is used instead of the value for amorphous carbon since 
significantly more dust is formed in these models. 
Even a small
increase of about 10\% in the intrinsic density of the dust material (as is the case from
Model l13dj10$\rho$199 to l13dj10$\rho$225)
results in a doubling of the degree of condensation and a
substantial increase in the outflow velocity, ${\langle u \rangle}$.

\begin{table}
\caption{Influence of the intrinsic density of the dust. 
         {\it Model parameters:} Same as Table\,2.} 
\begin{center}
\begin{tabular}{|l|c|c|c|c||c|c|c|}
\hline
Model& $L_{\star}$ & $T_{\star}$ & $\kappa_{\rm dust}$& $\rho_{\rm dust}$ & $\dot{M}$ & ${\langle u \rangle}$ & ${\langle {f_{\rm c}} \rangle}$  \\
 &  [$L_{\odot}$] & [K] & & g/cm$^{3}$ & [${\rm {M_{\odot}} / yr}$] & [km/s] & \\
\hline
l13dj10$\rho$199& 13000 & 2700& {\it J\"ager~1000} & 1.99 & $5.59 \cdot 10^{-6}$ & 14.8 & 0.05  \\
l13dj10$\rho$225&13000 & 2700& {\it J\"ager1000} & 2.25 & $7.26 \cdot 10^{-6}$ & 20.8 & 0.11  \\
l13drou$\rho$185&13000 & 2700& {\it  Rouleau} & 1.85 &  $4.87 \cdot 10^{-6}$ & 7.38 & 0.10   \\
l13drou$\rho$225& 13000 & 2700& {\it Rouleau} & 2.25 &  $8.24 \cdot 10^{-6}$ & 18.1 & 0.31  \\
l13dj04$\rho$144&13000 & 2700& {\it J\"ager400} & 1.44  & - & - & - \\
l13dj04$\rho$225& 13000 & 2700& {\it J\"ager400} & 2.25  & $2.10 \cdot 10^{-8}$ & 1.38 & 0.13 \\
\hline
\end{tabular}
\end{center}
\end{table}

The influence on the mass loss rate is very sensitive to  
the discrepancy between the value of the intrinsic 
density of the amorphous carbon and the value for graphite. 
For the models l13drou$\rho$185 and l13drou$\rho$225 where
the only difference is a 20\% difference in the value used for the intrinsic
density, the estimated mass loss rate differ almost by a factor of 2.
For the models using the {\it J\"ager~400} material data
(l13dj04$\rho$144 to l13dj04$\rho$225), where the difference is almost 
40\%, using the true material value instead of  
the higher value for graphite result in a model that will not develop 
a wind at all.

\subsection{Sticking coefficient} 

The sticking coefficient 
(also called the reaction efficiency
factor) $\alpha$ is not definitely know as long as we do
not know explicitly the sequence of chemical reactions
responsible for the dust formation.
To isolate the importance of how uncertainties in the
value of $\alpha$ will influence the results of the
models we have varied this parameter in otherwise identical
models. Choosing a value for the sticking coefficient of one 
increases the degree of condensation and the outflow velocity by
about a factor of 2 compared to when a value of 0.3 is chosen.
Dust formation occurs beyond $\approx 2 R_{\star}$ 
independent of the value assumed for the sticking coefficient
since the condensation threshold is determined mainly by the temperature.  With the low
choice of sticking coefficient ($0.3$) only about 10\% of the condensable
carbon material present in the gas actually condense into grains, while for the
much more favorable choice of sticking coefficient ($1$) twice as much
material condense into grains. A complete condensation of carbon grains is
prevented by the rapid velocity increase after the onset of avalanche nucleation
and the subsequent rapid dilution.

\begin{table}
\caption{Influence of the sticking coefficient for the dust formation.  {\it Model parameters:} 
       Same as Table\,2, except for the sticking coefficient 
       $\alpha_{C}$ $\alpha_{C_{2}}$ $\alpha_{C_{2}H}$  $\alpha_{C_{2}H_{2}}$, the
       dust opacity $\kappa_{\rm dust}$ = {\it Rouleau}, and 
       the intrinsic dust density $\rho_{\rm dust} = 1.85$ g/cm$^{3}$.} 
\begin{center}
 \begin{tabular}{|l|c|c|c|c|c|c||c|c|c|c|}
 \hline
Model& $L_{\star}$ & $T_{\star}$  & $\alpha_{C}$ & $\alpha_{C_{2}}$ & $\alpha_{C_{2}H}$ & $\alpha_{C_{2}H_{2}}$ & $\dot{M}$ & ${\langle u \rangle}$ & ${\langle {f_{\rm c}} \rangle}$  \\
 & [$L_{\odot}$] & [K] & & & & & [${\rm {M_{\odot}} / yr}$] & [km/s] & \\
 \hline
l13$\alpha$03& 13000 & 2700 & 0.37 & 0.34 & 0.34  & 0.34   & $4.87 \cdot 10^{-6}$ & 7.38 & 0.10 \\
l13$\alpha$02&  13000 & 2700 & 0.20 & 0.20 & 0.20 & 0.20 & $3.38 \cdot 10^{-6}$ & 3.87 & 0.09 \\
l13$\alpha$05& 13000 & 2700 &  0.50 & 0.50 & 0.50 & 0.50 & $5.75 \cdot 10^{-6}$ & 10.5 & 0.12 \\
l13$\alpha$10& 13000 & 2700 &  1.00 & 1.00 & 1.00 & 1.00 & $7.02 \cdot 10^{-6}$ & 16.5 & 0.22  \\
l10$\alpha$03&  10000 & 2600 &  0.37 & 0.34 & 0.34 & 0.34 & $2.26 \cdot 10^{-6}$ & 3.64 & 0.12 \\
l10$\alpha$10&  10000 & 2600 & 1.00 & 1.00 & 1.00 & 1.00 & $6.95 \cdot 10^{-6}$ & 12.3 & 0.23 \\
\hline
\end{tabular}
\end{center}
\end{table}

\subsection{Surface tension} 

The surface tension $\sigma_{\rm dust}$ of amorphous carbon is not know from
laboratory experiments, it has therefore become a custom to use
the values for graphite e.g.\ given by Tabak et al.\ (1975).
One problem however is that there are huge variations in the
values determined due to the anisotropy with orientation of graphite.

The surface tension enters into the surface contribution to the 
free energy, $\Delta F$,
calculated on the basis of classical thermodynamics,
see Gail et al.\ (1984) for details. 

As already demonstrated by Tabak et al.\ (1975)
then variating the value of the surface tension produces an
enormous change in the nucleation rate. To determine how significant
the prescribed value for the surface tension that we use in the
models is for the results we have varied the value around the
value of 1400~erg/cm$^{2}$ which were used in all previous models,
see Table\,5. Altering the value of the
surface tension of the forming dust grains by 28\% around the value
of graphite will make the difference between obtaining a mass loss
or not for the model. Also it has a substantial influence on how
much of the available material in the circumstellar envelope
will condense into dust grains. For a comparison the measured surface
tension for other materials are: Fe ($\sigma_{\rm dust} = 1400$~erg/cm$^{2}$), MgS
($\sigma_{\rm dust} = 800$~erg/cm$^{2}$)  and SiO ($\sigma_{\rm dust} = 500$~erg/cm$^{2}$) (Gail \& Sedlmayr 1986).

\begin{table}
\caption{Influence of the surface tension for the dust formation. 
     {\it Model parameters:} Same as Table\,2
     but also including the surface tension $\sigma_{\rm dust}$ and with the 
     dust opacity $\kappa_{\rm dust}$ = {\it Rouleau},
     and  the intrinsic dust density $\rho_{\rm dust} = 1.85$ g/cm$^{3}$.} 
\begin{center}
 \begin{tabular}{|l|c|c|c||c|c|c|}
 \hline
Model& $L_{\star}$ & $T_{\star}$  & $\sigma_{\rm dust}$ & $\dot{M}$ & ${\langle u \rangle}$ & ${\langle {f_{\rm c}} \rangle}$  \\
  &  [$L_{\odot}$] & [K] & [erg/cm$^{2}$] & [${\rm {M_{\odot}} / yr}$] & [km/s] &  \\
 \hline
l13drou$\rho$185$\sigma$10& 13000 & 2700 & 1000 & $8.54 \cdot 10^{-6}$ & 3.52 & 0.76 \\
l13drou$\rho$185$\sigma$14& 13000 & 2700 &1400 & $4.87 \cdot 10^{-6}$ & 7.38 & 0.10  \\
l13drou$\rho$185$\sigma$18& 13000 & 2700 & 1800 &  - &  - & - \\
\hline
\end{tabular}
\end{center}
\end{table}

\section{Conclusions}

We have investigated in detail how the predicted wind properties
of the carbon-rich AGB models are influenced by the choice of
micro-physical dust parameters.

For the mass loss models it is important to know
how the uncertainty in the chosen dust parameters
affects the obtained results.
The choice of the micro-physical parameters can change the
predictions for the mean outflow velocity of the gas and dust by a factor
of 4, the predicted degree of dust condensation by a factor of 10 and the
predicted mass loss by a factor of 2. In the transition between models
with and without mass-loss the choice of micro-physical parameters
is vital for whether a particular set of stellar
parameters will give rise to a dust-driven mass loss or not.

\acknowledgments
ACA gratefully acknowledges financial support from the Carlsberg  
Foundation. This work was supported by NorFA, the Royal Swedish  
Academy of Science and the Swedish Research Council.

%

\end{document}